\documentclass[a4paper, 11pt]{article}
\usepackage[margin=2.5cm]{geometry}
\usepackage[bibstyle=apa]{biblatex}
\usepackage{authblk}
\usepackage{booktabs}
\usepackage{graphicx}
\usepackage{bm}
\usepackage{subcaption}
\usepackage{amsmath}
\usepackage{hyperref}
\title{The Rest is Silence: Leveraging Unseen Species Models \\ for Computational Musicology}

\author[1]{Fabian C. Moss}
\author[2]{Jan Haji\v{c} jr.}
\author[3]{Adrian Nachtwey}
\author[4]{Laurent Pugin}

\affil[1]{Institut für Musikforschung, Julius-Maximilians-Universität Würzburg, Würzburg, Germany}
\affil[2]{Institute of Formal and Applied Linguistics, Charles University, Prague, Czech Republic}
\affil[3]{KreativInstitut.OWL, Paderborn University, Paderborn, Germany}
\affil[4]{RISM Digital Center, Bern, Switzerland}

\begin{document}

\maketitle

\begin{abstract}
    \noindent
    For many decades, musicologists have engaged in creating large databases serving different purposes
    for musicological research and scholarship. With the rise of fields like music information retrieval 
    and digital musicology, there is now a constant and growing influx of musicologically relevant datasets and corpora.
    In historical or observational settings, however, these datasets are necessarily incomplete,
    and the true extent of a collection of interest remains unknown --- silent.
    Here, we apply, for the first time, so-called Unseen Species models (USMs) from ecology to areas of musicological activity.
    After introducing the models formally, we show in four case studies how USMs can be applied to musicological data to
    address quantitative questions like: 
    How many composers are we missing in RISM? 
    What percentage of medieval sources of Gregorian chant have we already cataloged?
    How many differences in music prints do we expect to find between editions? 
    How large is the coverage of songs from genres of a folk music tradition?
    And, finally, how close are we in estimating the size of the harmonic vocabulary of a large number of composers?
\end{abstract}

\section{Introduction}

Many research questions in the Computational Humanities rely on distributional data about
cultural objects and artefacts, often gathered in observational rather than controlled experimental studies.
Found distributions of these objects are thus heavily shaped by uncertainties associated with historical transmission processes, including missingness (e.g. because something lies hidden in some basement) or loss (e.g. because a library burnt down). What's worse, if there is no external record about a missing or lost item, there is no way of knowing that it had ever existed. In order to understand the representativeness of observed samples in the humanities, it is thus of great interest to be able to gauge how much we should expect to be missing. 

This problem structurally resembles similar issues in species ecology, where researchers need to estimate the number of species from a limited set of incomplete samples. These belong to the class of so-called Unseen Species models (USMs), and they have recently been employed in cultural contexts as well: specifically in computational literary studies, both for estimating the true size of Shakespeare's vocabulary \cite{Efron1976_EstimatingNumberUnseen}, and recently in a comparative study of medieval chivalric epics across different European cultures, with focus on loss rates \cite{kestemont2022unseen}. However, in musicology, the potential of these models to give a better grasp of the unknown, to guide quantitative musicological research, and to provide new guidance for data collection efforts is far from realised.

In this contribution, we probe the usefulness of USMs for musicology in an array of four different case studies using large public sets of musicological data. We aim to demonstrate what kinds of questions can be addressed when applying USMs in music research, and to provoke discussions about the strengths and limits of this approach. 

To that end, we first introduce the methodology in Section~\ref{sec:methods}, and apply it to four musicological case studies in Section~\ref{sec:compmus}, namely the RISM and Cantus databases (Section~\ref{sec:RISM_Cantus}), a dataset of differences between 19th-century music prints (Section~\ref{sec:differences}), a dataset of folks music sessions (Section~\ref{sec:sessions}), and, finally, a large corpus of harmonic annotations in pieces from different composers (Section~\ref{sec:vocabulary}). We discuss our results in Section~\ref{sec:discussion} and conclude with some remarks of their implications and potential for (computational) musicology (Section~\ref{sec:conclusion}).

\section{Methods: Unseen species models in the Computational Humanities}\label{sec:methods}

Translating musicological data collection efforts to the context of unseen-species models means asking the question:

\begin{itemize}
    \item How many distinct cultural units (unknown species) did we not observe yet?
\end{itemize}

\noindent
This question is critical to the representativeness of musicological data. For example, the Cantus database has ``only'' indexed a few hundred of the tens of thousands of extant manuscripts of Gregorian chant \cite{helsen2014omr} --- but chant is supposedly a highly conserved, stable tradition. How well is the entire Gregorian repertoire covered by these sources? 

\paragraph{Modeling the abundance of species --- cultural units.}
In virtually any kind of collection of cultural objects, some units are highly abundant (very common; e.g., the same composition occurs in many manuscripts, the same song is known across an entire region, the same musical patterns occurs over and over in a composer's work), while others are rare, possibly occurring only once or twice. The (unknown) probability to encounter an instance of a certain unit --- in other words, observing a specimen of a species --- is thus proportional to the number of times that unit occurs in the whole tradition (also unknown, in general). This is called the \textit{relative abundance} (or relative frequency) $p_i$ of a unit $i$, and probability theory ensures that $p_i\geq 0$ and $\sum_i p_i = 1$, for $i=1,\dots, S$, where $S$ stands for the total number of cultural species (the quantity of interest here).\footnote{Relative abundances have been generalised to cover also the probability of species being detected in the first place \cite{chao2017deciphering}.}
For an observed collection of $n$ cultural units $i$ with abundances $X_i$, it holds that
\begin{align}
    \sum_{i = 1 \dots S} X_i = n.
\end{align} 

For some --- possibly many or even most --- species, $X_i=0$, i.e. they are not observed, their abundance in the sample is zero. While they were at some point present, written down or perhaps sung by someone, they don't figure in the particular dataset/catalog/songbook observed.
The \textit{absolute frequency} of species with a particular abundance $r$ is defined as:
\begin{align}
    f_r = \left|\{X_i\mid X_i=r\}\right|.
\end{align}
This implies that $f_1$ is the number of species observed only once (singletons)\footnote{In the context of corpus studies in natural language processing (NLP), singletons are sometimes called \textit{hapax legomena} (Greek for `read only once')~\cite{Manning2003_FoundationsStatisticalNatural}.} and $f_2$ is the number of species observed twice (doubletons). With $f_0$ we denote the (unknown) number of species that was not observed. It logically follows that the total number of distinct species observed in the sample, $S_{\mathrm{obs}}$, is given by 
\begin{align}
    S_{\mathrm{obs}} = \sum_{r = 1 \dots \infty} f_r. \label{eq:Sobs}
\end{align}
This, finally, allows us to arrive at defining our quantity of interest, $S$, which is the true size of the musical tradition (for which we are only looking at a limited sample) measured in terms of the number of distinct musical items (songs, manuscripts, harmonies, etc.). It is simply the number of items observed plus the number of items not observed:
\begin{align}
 S = S_{\mathrm{obs}} + f_0.  \label{eq:4}
\end{align}
Since $S_{\mathrm{obs}}$ is known (Eq.~\ref{eq:Sobs}) and the true value of $f_0$ cannot be known, finding $S$ relies on realistic estimates for $f_0$ based on limited samples. This is where Unseen Species models from ecology come into play. 

\paragraph{Modeling species incidence.}
For some musicological scenarios, it seems more appropriate to track \textit{incidence}, i.e., the presence or absence of some musical unit in a sample, instead of counting how many times each individual species was observed. 
Incidence-based models look at $m$ different samples and re-define $f_r$ to represent the number of species observed in $r$ samples (instead of $r$ times in a single sample in the case of species abundance). 
That is, we only care about \textit{whether} a cultural unit has appeared in a sample. 

Decisions such as choosing between abundance and incidence are the responsibility of the experiment designer(s) and depend on factors that can better be addressed by theory than by empiricism~\cite{deffnerBridgingTheoryData2024}. 

\paragraph{Chao estimators.}
Popular estimators for species abundance and incidence are the Chao estimators \cite{chao1984estimator}, which have already been applied in the computational humanities~\cite{karsdorpIntroducingFunctionalDiversity2022,kestemont2022unseen}. Among those, ``Chao1''\footnote{%
Other common methods are the Abundance Coverage Estimator (ACE)~\cite{chao2004ace}, Jackknife~\cite{smith1984jackknife}, and Good-Toulmin estimators~\cite{orlitsky2016optimal,hao2020multiplicity}. Here, we opt for using Chao estimators because a) they provide a conservative lower bound \cite{kestemont2022unseen}; b) because they have already been used in other applications of the unseen species model to the humanities; and c) because they are straightforward to compute; and d) because their interpretation naturally flows from their formal logic. We use the implementation of unseen species estimators from the \texttt{copia} Python library \cite{karsdorpIntroducingFunctionalDiversity2022}.
} 
 is one particular way of estimating $S$ from the relative frequency counts $f_r, r > 0$.
It is formally defined as:
\begin{align}
    S = S_{\mathrm{obs}} + \frac{f_1^2}{2 \cdot f_2}, \label{eq:5}
\end{align}
that is, it estimates the number of species yet unseen from the numbers of species observed only once or twice.\footnote{In cases where both abundance and incidence is studied, $f_r$ for incidence is commonly denoted by $Q_r$, and the incidence-based estimator is called ``Chao2''~\cite{chao1984estimator,chao2017deciphering}.}
From Equations~\ref{eq:4} and~\ref{eq:5} follows that the number of unseen species is given by 
\begin{align}
    f_0 = \frac{f_1^2}{2 \cdot f_2} = S - S_{\mathrm{obs}},
\end{align}
and we define \textit{species coverage} (the overall ratio of species already observed) as
\begin{align}
    \hat{c}=\frac{S_{\mathrm{obs}}}{S}.
\end{align}

As mentioned above, the estimator is based solely on the count of singletons ($f_1$) and doubletons ($f_2$). 
The basic intuition behind this assumption is that distributions of counts 
usually have
a `long tail': few items occur very frequently and many items occur rarely~\cite{clausetPowerLawDistributionsEmpirical2009}, and the probability mass available for yet unseen species should follow from the length of the tail.\footnote{A constructive proof derived from Good-Turing smoothing \cite{good1953smoothing,gale1995good} is provided in \cite{chao2017deciphering}.} 

Crucially, this estimator is non-parametric \cite{chao1984estimator}; that is, it it can be used regardless of the underlying distribution of relative abundances or incidences $p_i$. This is its major advantage because this distribution is usually unknown and allows for the application of Chao1 across a wide range of research areas.

The Chao estimates for $f_0$ are lower bounds \cite{chao2017deciphering}: they provide the \textit{minimum} expected number of unseen species (we could still be missing more). Consequently, the estimated species coverage constitutes an \textit{upper} bound: a value of 0.5 means that we have observed \textit{at most} half of all the musical species in some collection or repertoire. 

\section{Applications for Computational Musicology}\label{sec:compmus}

A crucial step in applying Unseen Species Models from ecology to cultural contexts in the humanities in general and to musicology in particular is to draw convincing analogies of what the concept of species corresponds to.
Here, we present four case studies relevant for different branches of musicological research. We start by looking at unseen composers and repertoire in large collections of music sources (historical musicology), move on to differences in music prints (music philology), followed by analyzing sessions of folk musicians (ethnomusicology). We conclude by applying the model to harmonic vocabularies in different repertoires (music theory).

For each case study, we draw an analogy between biological species and some music(ologic)al entity. 
This enables us to apply Chao1 and estimate how many species are missing from this particular context.
We also calculate quantities more commonly used in the digital and computational humanities, 
namely the numbers of types and tokens --- corresponding to incidence and abundance data, respectively --- 
as well as the ratio between them, the so-called type-token ratio (TTR).
Intuitively, TTR and Chao1 should stand in a reciprocal relation: a TTR of 1 means that there are as many tokens as there are types, which implies that each type occurs exactly once. On the other hand, a low TTR (which can approach, but not reach zero), means that at least some types occur many times, potentially leading to a longer tail and thus to a higher estimate of yet unseen species. We thus expect TTR and Chao1 to be negatively correlated. But since the two quantities are based on very differing assumptions and calculations, one should not expect this correlation to be particularly strong.

\subsection{Case Study 1: Databases and archives}
\label{sec:RISM_Cantus}

Empirical conclusions drawn about a musical tradition from a database or corpus rely on the representativeness of the given data source. While one cannot know what is not represented in a database (because the effort could then just be spent better by adding the given items to the database in question!), we can use the Unseen Species models to estimate \textit{how much} of whatever entity we define as the ``species'' is not covered by the data source. We can thus quantify how much ``cultural diversity'' has not yet been.

We present here reports from two of the largest musicological databases: RISM, by far the largest database of musical sources, and the Cantus database of Gregorian chant.

\subsubsection{RISM: Counting composers}

How many composers were active in Europe since the Renaissance? 
How many composers are we still to discover whose works lie undetected on some attic or archive shelf?

There is no better database to answer such questions than the \textit{Répertoire International des Sources Musicales} (RISM),\footnote{\url{https://rism.online/}} a database of more than 1,500,000 musical sources assembled across nearly 3,000 holding institutions.
In this setting, each composer (which are identified by a RISM authority record) can be thought of as a species, and the presence of a composer's work in a source catalogued in RISM is then an observation of that species, a concrete instantiation of that composer's work.
For a given institution, the composer observation count is the number of sources held by that institution in which the composer appears.

The resulting dataset contains records for 48,524 composers\footnote{In the usual sense of the term: persons who are identified as authors of written musical works. We do not consider phenomena such as recording folk musics, or rather: we accept editorial decisions made by those who catalogued records in RISM.} with works observed across 2,933 holding institutions, with a total of 2,009,343 observations of composers appearing in sources.
A composer is a singleton if they are observed in only one source in a single institution, and a doubleton if observed exactly in two sources, either in the same institution or across two of them.

Aggregated over the entire dataset, the Chao1 estimate gives us a lower bound of 78,432 species --- composers. With 48,524 composers observed, that implies an $f_0$ of 29,908 unobserved composers and a coverage upper bound of 0.619 --- we have so far recorded in the RISM database at most some 62\% of all composers, indicating that there might be plenty of musical diversity to discover.

\begin{figure*}[t]
  \centering
  \begin{subfigure}[t]{0.49\textwidth}
    \includegraphics[width=\linewidth]{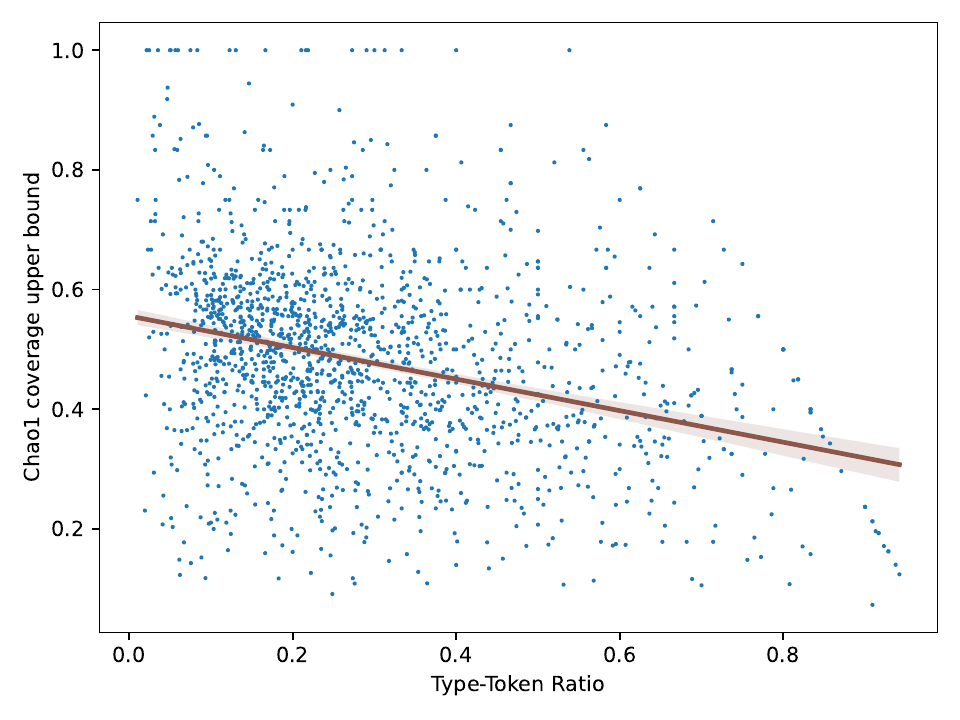}
    \caption{Type-Token Ratio (TTR) per institution in RISM plotted against the Chao1 coverage upper bound. Linear regression shows a best fit at slope $-0.25$, with the p-value for non-correlation (zero slope) below $10^{-30}$, but the variance of the linear estimate is large.}
    \label{fig:rism:ttr}
  \end{subfigure}
  \hfill
  \begin{subfigure}[t]{0.49\textwidth}
    \includegraphics[width=\linewidth]{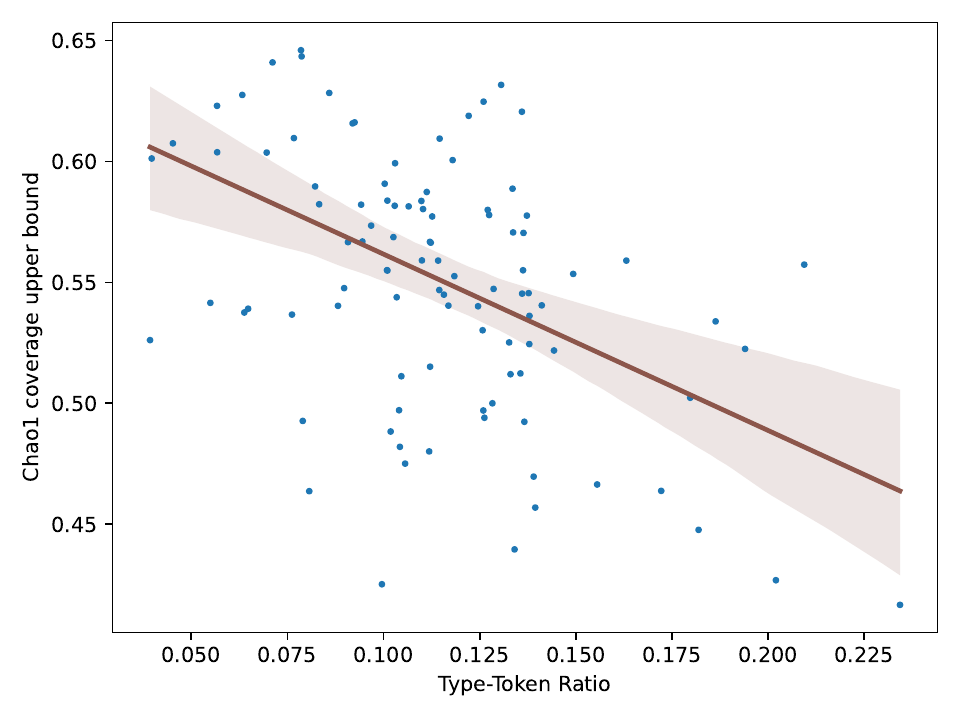}
    \caption{Type-Token Ratio (TTR) in relation to Chao1 coverage upper bound on the 100 largest institutions in RISM, to counteract the effect small data has on biasing TTR higher than in the sampled population. In this case, linear regression shows a best fit at slope $-0.69$, with the p-value for non-correlation below $10^{-5}$.}
    \label{fig:rism:ttr100}
  \end{subfigure}
  \caption{RISM results for the relationship between the Chao1 coverage upper bound and linear proxies for diversity: the Type-Token Ratio (TTR). Note that the upper right triangle plot (a) is empty: this is because at very high TTRs, the Chao1 coverage cannot be very high: even with the most uniform distribution possible, at TTR > 0.5 there will always be at least one singleton contributing to $f_1$, so coverage upper bound cannot be 1.0, and at TTR close to 1.0, nearly all tokens will contribute to $f_1$ and coverage upper bound will thus approach 0 (lower right corner). However, the correlation between TTR and Chao1 coverage is not caused by this: when we restrict ourselves to institutions where TTR < 0.6 and coverage < 0.8, where empirically this effect does not reach, we still get Pearson's $r$ of $-0.21$ and $p < 10^{-15}$.}
  \label{fig:rism:correlations}
\end{figure*}

If we aggregate results only over the 10 largest institutions, each of which holds more than 20,000 composer records, we get an estimated $S = 32,989$ total composers with $S_{\text{obs}} = 20,778$ composers observed, with a similar coverage of $0.630$. For the 100 largest institutions, in turn, with $S_{\text{obs}} = 34,090$, we obtain $S = 53,561$ and coverage 0.635. We interpret this to indicate as sampling error: the combined largest music libraries are still not sampling the same space of composers with extant works as all the holding institutions, including the smaller ones. Otherwise we should see a similar estimate of total composers around 80,000 as for the complete dataset, with the corresponding coverage upper bound of approx. 0.26. This implies that smaller institutions play an important role in documenting the overall diversity of composers. Only when we aggregate data over the top 600 institutions do we get the Chao1 estimate of at least 70,000 composers. 
We compute the TTR and Chao1 coverage $S_{\text{obs}} / S$ for each RISM institutions, and from these value pairs we measure how these metrics are related. While there is some relationship between TTR and Chao1 coverage, it is weak, and has a very high variance. For the 100 largest institutions, Pearson's $r = -0.46$ and non-correlation can be rejected (p-value for non-correlation using \texttt{scipy.stats.linreg}: $<10^{-5}$), and the same holds for all institutions, though the relationship is even weaker ($r = -0.295, p < 10^{-30}$). The relationship between TTR and Chao1 is shown in \autoref{fig:rism:correlations}.

\subsubsection{Cantus: ``biodiversity'' of Gregorian chant}

The Cantus database\footnote{\url{https://cantusdatabase.org/}} is a large-scale project for cataloguing Gregorian chant that has been running since the mid-1980s \cite{lacoste2012cantus, lacoste2022cantus}. It primary mechanism is the Cantus ID, which identifies instances of the same chant --- element of Gregorian repertoire --- across multiple manuscripts. Cataloguing a manuscript means primarily assigning Cantus IDs to all chants occurring therein. Chant manuscripts often have more than 1,500 chants, so cataloguing even a single manuscript requires considerable effort.

Gregorian chant was --- is --- an immense tradition. It was one of the markers of cultural identity of medieval Latin Europe \cite{hiley1993western}, and later of its colonial reach. 
Despite the initial Carolingian project of Gregorian chant as a shared and tightly controlled expression of a common identity \cite{hiley1993western}[p.514--523],
over the centuries and across Latin Europe, repertoire choices diversified greatly, to the extent that one can often identify the provenance of a manuscript directly from the repertoire choices. This diversity, combined with the scale of chant, permits one to think of the Gregorian tradition in terms of ecology, and ask: to what extent has the Cantus database covered the existing ``biodiversity'' of chant?

\begin{table}[t]
	\centering
	\begin{tabular}{lrrrrrrrr}
	\toprule
\textbf{Genre} & \textbf{Cantus IDs} & \textbf{Mss.} & \textbf{Tokens} & \textbf{TTR} & \textbf{STR} & \bm{$f_1$} & \bm{$f_2$} & \textbf{Coverage}\\
	\midrule
\textbf{A}	& 11158	& 231	& 205409	& 0.054	& 0.021	& 4714	& 1542	& 0.569 \\
\textbf{R}	& 5099	& 213	& 102443	& 0.050	& 0.042	& 2151	& 714	& 0.553 \\
\textbf{V}	& 8163	& 214	& 94482	& 0.086	& 0.026	& 3919	& 1068	& 0.502 \\
\textbf{W}	& 926	& 185	& 35594	& 0.026	& 0.200	& 292	& 127	& 0.679 \\
\textbf{I}	& 600	& 181	& 10086	& 0.059	& 0.302	& 250	& 106	& 0.596 \\
	\midrule
\textbf{In}	& 207	& 50	& 2025	& 0.102	& 0.242	& 41	& 5	& 0.573 \\
\textbf{InV}	& 286	& 32	& 1294	& 0.221	& 0.112	& 82	& 32	& 0.745 \\
\textbf{Gr}	& 154	& 90	& 2411	& 0.064	& 0.584	& 28	& 9	& 0.733 \\
\textbf{GrV}	& 207	& 68	& 1677	& 0.123	& 0.329	& 53	& 11	& 0.666 \\
\textbf{Al}	& 405	& 73	& 2382	& 0.170	& 0.180	& 159	& 62	& 0.624 \\
\textbf{AlV}	& 38	& 29	& 189	& 0.201	& 0.763	& 24	& 3	& 0.197 \\
\textbf{Of}	& 158	& 43	& 2209	& 0.072	& 0.272	& 25	& 17	& 0.810 \\
\textbf{OfV}	& 263	& 13	& 764	& 0.344	& 0.049	& 44	& 39	& 0.901 \\
\textbf{Cm}	& 198	& 42	& 2155	& 0.092	& 0.212	& 27	& 8	& 0.731 \\
\textbf{CmV}	& 154	& 4	& 253	& 0.609	& 0.026	& 135	& 16	& 0.183 \\
\textbf{Tc}	& 47	& 21	& 294	& 0.160	& 0.447	& 10	& 6	& 0.797 \\
\textbf{TcV}	& 203	& 21	& 858	& 0.237	& 0.103	& 44	& 27	& 0.846 \\
\bottomrule
	\end{tabular}
	\caption{Unseen species estimates for CantusCorpus v0.2 data, split by genre. We report the number of distinct chants (Cantus IDs) for each genre, the number of manuscripts (because we are using incidence data: Mss.), the number of tokens (in this case: total number of chants catalogued), the Type-Token Ratio (which is computed from chant counts), additionally the Sample-Type Ratio, which is an analogy for TTR for incidence data (ratio of the sample count to species count), the singleton and doubleton counts used for Chao1 estimation, and the resulting Chao1 upper bound on coverage. Note how the coverage varies between genres (especially those for the Mass Propers).}
	\label{tab:cantuscorpus:results}
\end{table}

In this abstraction, the Cantus IDs serve as species, and manuscripts serve as samples. We ask: how much of chant repertoire has been catalogued, and how much remains to be discovered?

Chant repertoire is categorised according to \textit{genre}, its function in liturgy. For example: antiphons are short and simple chants that are sung before and after psalms; responsories are longer and more ornate chants that are sung between blocks of psalms (with the antiphons). The most ornate and virtuosic chant genre is probably the gradual and its verses, sung in Mass. Individual chant genres had a varied history as liturgy developed: for instance, the Offertory verses (another genre sung in Mass) fell out of use after the 13th century \cite[p.121]{hiley1993western}. It therefore makes sense to quantify the (in)completeness of the Cantus database according to the individual main genres.

In this case study, we apply an incidence-based approach over abundance. Instead of counting how many times each chant appeared in the dataset, we count its presence or absence in sources. Preferring incidence follows naturally from the structure of chant data. Each written source acts as a sample from one site: a particular ecclesiastical community.
A chant recorded in only one source, even if used twice or more times, is still considered a singleton and contributes to $f_1$. 
We believe this to be more appropriate: a chant being used in more than one liturgical position in a certain church should not necessarily imply the particular chant would be more likely to be used in other churches.

We use the CantusCorpus~v0.2 dataset \cite{cornelissen2020studying}, which is a dataset derived from the Cantus database that is most widely used for computational chant research \cite{cornelissen2020mode,helsen2021sticky,lanz2023,lanz2025modality}.

We report the Chao1 upper bounds on coverage for individual genres on CantusCorpus~v2.0 in \autoref{tab:cantuscorpus:results}. The genres of chant for one type of liturgy, the Divine Office (upper section of \autoref{tab:cantuscorpus:results}), exhibit overall lower maximum coverage than the chants for Mass (lower section of \autoref{tab:cantuscorpus:results}) This is quantitative confirmation that Mass repertoire was more stable, possibly because the Mass is a public liturgy while the Divine Office is primarily a private prayer for the clergy and thus the processes for adapting the rite may have been easier and more localised (though still with bureaucracy involved \cite{hallas2021offices}), though interestingly the Introit genre, which starts the Mass, seems to be covered less well.

\begin{figure*}
  \centering
  \begin{subfigure}[t]{0.49\textwidth}
    \includegraphics[width=\linewidth]{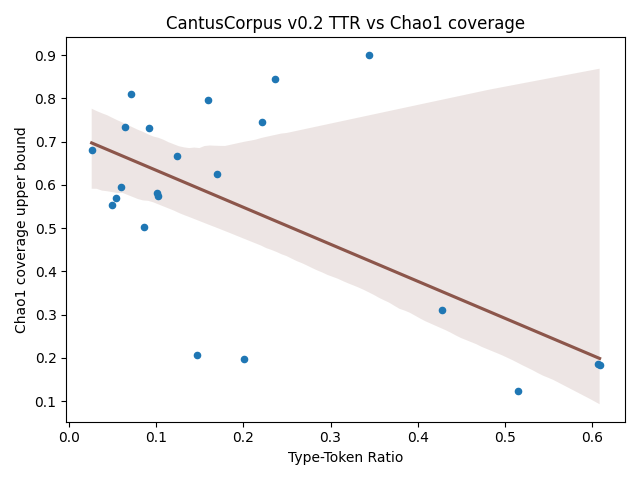}
    \caption{Type-Token Ratio (TTR) per chant genre on CantusCorpus v0.2 plotted against the Chao1 coverage upper bound. Linear regression shows a best fit at slope $-0.78$, with the p-value for non-correlation (zero slope) at $0.005$, but the variance of the linear estimate is large.}
    \label{fig:cantus:ttr}
  \end{subfigure}
  \hfill
  \begin{subfigure}[t]{0.49\textwidth}
    \includegraphics[width=\linewidth]{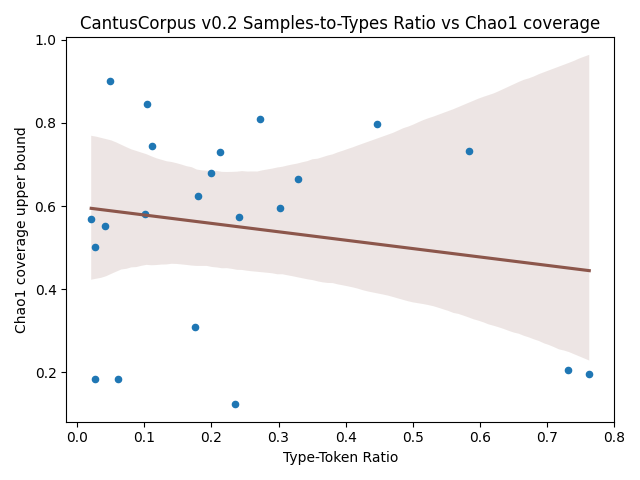}
    \caption{Sample-Type Ratio, an analogy of TTR for incidence-based data. In this case, linear regression shows a best fit at slope $-0.20$, with the p-value for non-correlation at $0.42$.}
    \label{fig:cantus:str}
  \end{subfigure}
  \caption{CantusCorpus results for the relationship between the Chao1 coverage upper bound and linear proxies for diversity: the Type-Token Ratio (TTR), and its analogy for incidence data, the Sample-Type Ratio (STR). While the TTR has a correlation of -0.77 and thus non-correlation can be rejected (p=0.005), predicting the Chao1 coverage upper bound still has a very large variance. STR is not correlated at all (p=0.42).}
  \label{fig:cantus:correlations}
\end{figure*}

Neither the Type-Token Ratio, nor its incidence-based analogy Sample-Type Ratio, are particularly good predictors of the Chao1 coverage upper bound. Linear regression on TTS vs. Chao1 coverage has a slope of $-0.77$ and non-correlation can be rejected ($p=0.005$), but the variance is large (see \autoref{fig:cantus:ttr}); for STR, non-correlation cannot be rejected ($p=0.42$; see \autoref{fig:cantus:str}).

\subsection{Case Study 2: Ontology of differences in music prints}\label{sec:differences}

In this case study, we examine visual/notational differences between six editions of Beethoven's Bagatelles Op. 33, Nos. 1—5. The editions we used are the first print by Bureau d'Arts [sic] et d'Industrie (c. 1803), Zulehner (c. 1808), André (c. 1825), Schott (c. 1826), Haslinger (c. 1845) and Breitkopf~\&~Härtel (1864).\footnote{All the editions can be found online at the Beethoven Haus Bonn (\url{https://tinyurl.com/BeethovenhausOp33}) except for the Breitkopf edition which can be found via the Petrucci Music Library (\url{https://imslp.org/wiki/7_Bagatelles,_Op.33_(Beethoven,_Ludwig_van)}.} 
The data for this analysis consists of files containing the results of comparisons of different MEI encodings of the six editions. 
Through comparisons using the Python tool \texttt{musicdiff}~\cite{Foscarin2019},\footnote{\url{https://github.com/gregchapman-dev/musicdiff.git}} we obtained the differences between each pair of encodings\footnote{Find the encodings here: \url{https://github.com/CorpusBeethoviensis/beethoven-diff-docker.git}} from which we extract the kinds and numbers of differences that occur.
The Bagatelles contain a total of $38,785$ differences, summed across the six editions of each of the seven Bagatelles (for a total of $15 x 7 = 105$ pairwise comparisons). There are $81$ different types of differences.

\begin{figure}[h]
\begin{subfigure}[t]{0.49\textwidth}
\includegraphics[width=\textwidth]{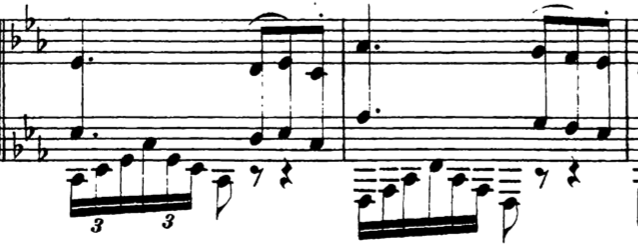}
\caption{Edition by Breitkopf \& Härtel.}
\end{subfigure}
\hfill
\begin{subfigure}[c]{0.45\textwidth}
\includegraphics[width=\textwidth]{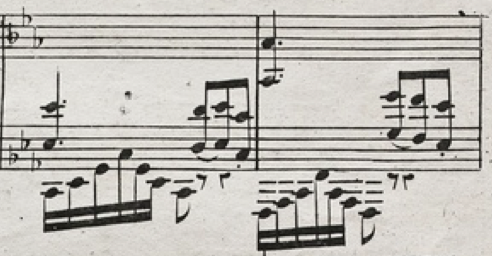}
\caption{Edition by Schott.}
\end{subfigure}
\caption{Example for differences between the editions of the 5th Bagatelle, bars 25 and 26, by Breitkopf \& Härtel and Schott. They differ in the placement of the right hand melody, of the articulations (slur and staccato) of this melody and the numbers to indicate triplets. Also, different symbols for quarter rests are used.}
\label{fig:editions}
\end{figure}

Some types differences are illustrated in Figure~\ref{fig:editions}. It shows two bars, 25 and 26, from the 5th Bagatelle in the editions of Breitkopf \& Härtel and Schott. In the edition on the left (Breitkopf) the melody in the right hand is split on the two systems with one note on the upper system and one on the lower one. On the right, the same melody is printed in the lower system (with one exception). There are also some less obvious differences, like the numbers to indicate triplets in the left hand in the first bar of the Breitkopf edition, which are omitted by Schott, or the placement of the slurs above two notes in the Breitkopf edition and below in the Schott edition. Also, the symbols to indicate a quarter rest are different.

For most of the transmission history of these works---and music in general---, prints played a crucial role, and they heavily influenced the way the broad interested public got to know them \cite{Lewis2024}. The process of copying music throughout this history affects the musical text by intentionally or accidentally introducing variants \cite{Grier1995}. Some printed notational variants do not affect the performance, like the one shown in Figure~\ref{fig:editions}. 
Despite the importance of prints for the historical reception of music, in musicological research, they are most often regarded as less important than other sources like manuscripts, even though the reception of music can heavily influence our own perception of these historical works today~\cite{Nachtwey2024}.

The species in this case study are the types of differences found in comparing all pairs of editions of each Bagatelle.
Thus, the question of unseen species in this case study is: how complete is this set of differences?
Compared to the previous case study, here the unseen species problem is not the representativeness of a sample of material, but a quantitative introspection of a constructed ontology. The implications of high coverage in such a context would be that very few new categories are likely missed, and therefore the given ontology can potentially be applied to a larger corpus as-is (e.g., via a machine learning model). 

The Chao1 estimate for the combined Bagatelles data is $S = 85$. With $S_{obs} = 81$, that means that the coverage of the differences ontology is nearly $95 \%$. This is an upper bound, so the true coverage may be lower, but it is unlikely that the ontology  still has significant blind spots.

How early could we estimate how many categories we \textit{should} first find before having a good chance of a near-complete ontology?
We run the estimation with 1000 sub-samples (without replacement) of different sizes $k$ and measure the average $S$ at a given $k$. At $k=1000$ selected out of the 38,785 differences, average $S$ is underestimated to be $67$, with $50$ categories observed; at $k = 5000$, $S = 76$ with average $S_{obs} = 66$, and at $k = 10,000$, somewhat above 25 \% of the total differences, we obtain $S = 80$, very close to the true $S_{obs} = 81$ categories.

If one estimates $S$ from all pairs of a single Bagatelle's editions, the estimates converge very quickly to the true number of difference categories for that particular Bagatelle. Using just 10~\% of the total differences from each Bagatelle's edition pairs, Chao1 underestimates the true number of categories by only $5.2 \%$ on average. However, the $S_{obs}$ for each complete Bagatelle never reaches more than $54$, so using a single Bagatelle to estimate the total $S$ is never going to enable reaching the true diversity of distinct editorial differences.
This is expected, as each Bagatelle contains specific musical material that uses only a subset of possible music notation patterns, and therefore certain types of editorial differences do not have a chance to appear (e.g., explicit vs. implicit triplets in a composition with no triplets). This result illustrates how sampling assumptions of Chao1 estimators might be violated, but conversely also how well the estimators work when this assumption holds.

\subsection{Case Study 3: Folk Music Sessions}\label{sec:sessions}

Musicians all over the world gather regularly to perform traditional Irish music~\cite{tolmiePlayingIrishMusic2013}. 
The platform \textit{The Session} tracks many if not most of these meetings,
and moreover hosts a rich database of tunes that its users have added,
including melodies of Irish tunes and their genres, such as Reel, Jig, Polka, Waltz, etc. 
Exports of the site's database are publicly available on GitHub.\footnote{\url{https://github.com/adactio/thesession-data}}
It has been shown that population size is an important factor for melodic variety~\cite{streetRolePopulationSize2022}. 
Specifically, while popular tunes recorded in the Sessions data set show higher variation of melodic complexity in their different settings, 
popularity is also strongly related to \textit{intermediate} complexity of tunes. Given this mainly performer-centered view, the tunes themselves have received somewhat less attention. 

The question that USMs can answer for this scenario is, 
given that sessions will continue to take place all over the world and 
that people record what was played on the website, how many tunes are likely to still be `out there,' 
either in an almost forgotten tunebook, or in someone's mind. 
Given the partition of this tradition into more or less well-defined genres, 
we can also ask whether there are between-genre differences
regarding the coverage of the repertoire. 

Table~\ref{tab:sessions_genres} shows an overview of all genres in the Session dataset and the numbers 
of pieces they contain.
It is sorted according to the coverage estimated with the Chao1 estimator. 
As can be seen, Marches have highest coverage of approx. 80.2\%., 
whereas the Mazurka coverage is relatively low at approx. 53.2\% 
--- not too surprising, given that Mazurkas are originally a Polish dance form. 
Taken together, the tunes recorded in the entire Session 
dataset are estimated to cover about 76.1\% of the entire Irish tunes repertoire. 
We can interpret this as showing that the tradition of Irish music sessions
is successful in preserving and transmitting its traditional repertoire. 
At the same time, some of the most common tune types, or genres, 
like Reels and Jigs have only moderate coverages of 77 and 73.8 percent, respectively. 
Thus, the differences in popularity of some genres clearly also affect their preservation. 

\begin{table}[]
    \centering
\begin{tabular}{lrrrrrr}
\toprule
\textbf{Genre} & \textbf{Types} &\textbf{Tokens} & \textbf{TTR} & \bm{$f_1$} & \bm{$f_2$} & \textbf{Coverage} \\
\midrule
March & 390 & 4212 & 0.093 & 110 & 63 & 0.802 \\
Slide & 269 & 5318 & 0.051 & 72 & 36 & 0.789 \\
Slip Jig & 430 & 10351 & 0.042 & 126 & 69 & 0.789 \\
Hornpipe & 749 & 12925 & 0.058 & 234 & 134 & 0.786 \\
Strathspey & 361 & 2504 & 0.144 & 112 & 59 & 0.773 \\
Barndance & 329 & 2698 & 0.122 & 114 & 67 & 0.772 \\
Reel & 4272 & 104131 & 0.041 & 1192 & 558 & 0.770 \\
Polka & 835 & 11857 & 0.070 & 271 & 145 & 0.767 \\
Three-Two & 101 & 516 & 0.196 & 34 & 17 & 0.748 \\
Waltz & 922 & 8104 & 0.114 & 329 & 166 & 0.739 \\
Jig & 2896 & 70826 & 0.041 & 931 & 421 & 0.738 \\
Mazurka & 109 & 888 & 0.123 & 48 & 12 & 0.532 \\
\midrule
\textbf{Total} & \textbf{11663} & \textbf{234330} & \textbf{0.050} & \textbf{3573} & \textbf{1747} & \textbf{0.761} \\
\bottomrule
\end{tabular}
    \caption{Repertoire coverage in different Irish folk genres represented in \textit{The Session} dataset. Pearson correlation of coverage and type-token ratio (TTR); $\rho=.28$ ($p=.35$).}
    \label{tab:sessions_genres}
\end{table}

\subsection{Case Study 4: Harmonic vocabularies}\label{sec:vocabulary}

In this case study, we use for the first time an estimator derived in the context of the Unseen Species problem for the question of the overall size of the harmonic vocabulary of Western tonal music. Since many years now, corpus studies in music theory have gained traction, and a variety of datasets have been created and made available for computational work. Chord vocabularies have been shown to follow both Zipf's~\cite{Moss2019_StatisticalCharacteristicsTonal,Zanette2006_ZipfLawCreation,Perotti2020_EmergenceZipfLaw,Manaris2005_ZipfLawMusic} and Heaps' laws~\cite{Moss2019_TransitionsTonalityModelbased,Serra-Peralta2021_HeapsLawVocabularya}, but these findings have not yet been extended 
to estimating what's still missing from empirical distributions of chords. 

The recently published \textit{Distant Listening Corpus} (DLC v2.3)~\cite{Hentschel2025_CorpusModularInfrastructure} consists currently of 40 sub-corpora, some of which had been previously published separately~\cite{neuwirthAnnotatedBeethovenCorpus2018,hentschelAnnotatedMozartSonatas2021b,hentschelAnnotatedCorpusTonal2023}. It encompasses 1,238 score encodings by 36 composers from the extended tonal tradition (c. 1550--1945). Each piece has been meticulously analyzed by music theory experts using harmonic labels conforming to an elaborate annotation scheme based on Roman numerals\footnote{\url{https://dcmlab.github.io/standards/}}. In total, there are 6,015 \textit{different} chord types, with a total abundance of 246,166 chords. 
Table~\ref{tab:composers} in Appendix~\ref{appdx:tables} gives an overview, showing the composers' names and their birth and death dates, the numbers of chord types and tokens as well as the type-token ratio (TTR) of their works contained in the DLC. Moreover, the numbers of singletons and doubletons, and the estimated coverage based on Chao1 are shown, too. Each row shows values for a particular composer, and the last row shows these values for the aggregated corpus.
Taking all DLC corpora together, the Chao1 estimator asserts that almost 70\% of the total harmonic vocabulary have been covered by this massive annotation effort, providing at the same time encouragement for its continuation. 
 
\begin{figure}[t]
    \centering
    \includegraphics[width=.8\linewidth]{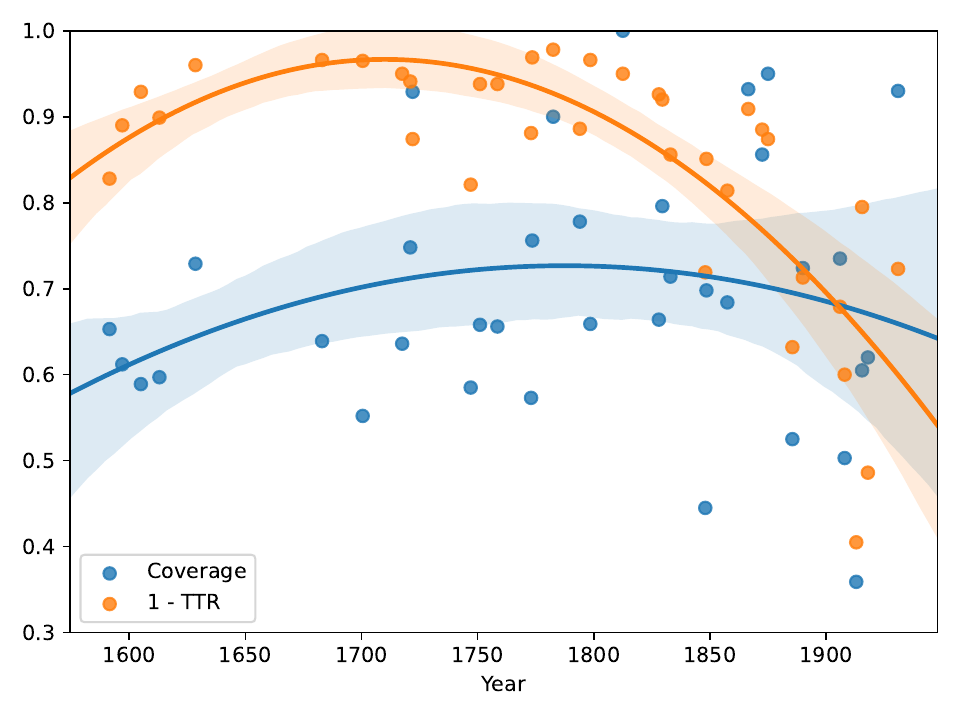}
    \caption{Vocabulary coverage (blue) and type-token ratio (TTR; orange) over time, with 2nd-order polynomial fit to the data points. Note that, for easier comparison, we show $1 - \text{TTR}$. Pearson correlation coefficient $\rho=.32$ ($p\approx.05$).}
    \label{fig:diachronic_composers}
\end{figure}

Figure~\ref{fig:diachronic_composers} compares the Chao1-based estimates of species coverage (blue) with the TTR values (orange) for each composer in the DLC.\footnote{DLC sub-corpora by the same composer were merged.} In order to facilitate the visual comparison, the figure shows $1 - \text{TTR}$. For both quantities, we added a quadratic regression; error bands represent a 95\% confidence interval based on bootstrap samples of the data.\footnote{See \url{https://seaborn.pydata.org/generated/seaborn.regplot.html} for details.} The error for the Chao1 estimates fluctuates more because the values are more widely dispersed, especially in the second half of the timeline. The Pearson correlation between the two sets of datapoints is $\rho=.32$ ($p\approx.05$), indicating a positive but weak association, as expected.

The interpretation of TTR and Chao1 in this context is not straightforward. 
The former shows the empirical fraction between the observed vocabulary size and chords used, but the number of chord tokens depends on many non-random factors, e.g. sonatas tend to be longer than Lieder, so a higher number of tokens may stem from a composer's preference for certain genres. Moreover, while the indicated curve of the TTR over time (orange line in Figure~\ref{fig:diachronic_composers}) \textit{could} be interpreted somehow to a real change of of the harmonic language over time, the curve fitted to the Chao1 estimates tells us rather something about where further encoding and annotation efforts should be directed. Apart from the general observation that many digital music corpora are heavily biased~\cite{Shea2024_DiversityMusicCorpus}, the coverage of the harmonic vocabularies of composers `at the fringes' of the represented timeline could be increased by sampling (i.e., encoding and annotating) more pieces from around that time---by the same or different composers.

\section{General Discussion}
\label{sec:discussion}

In this study we have applied the popular Chao1 model to estimate the abundance and incidence numbers of unseen species in range of specific cultural contexts that are commonplace in musicological scholarship. 
We have looked at two of the largest databases of musicological origin, RISM and Cantus, and estimated how many new composers and chants, respectively, we should still expect to encounter when continuing these cataloguing efforts. 
We have looked at the practice of 19th-century music prints and notational differences between them caused by to editorial intervention or pure chance, and have provided a principled answer to the question of how complete an ontology of these differences is.
In the domain of music performance, we have analyzed data about Irish folk music sessions and the repertoire coverage between different sub-genres. 
Addressing questions of music theory, we have looked at the size of analytical harmonic vocabularies of a great number of composers from the Renaissance onward. 

What have we learned from all of this? First of all that recognizing structural similarities between vastly different fields enables the transfer of methods and can lead to opening up entirely new avenues of research. 
Second, using the Chao1 estimators is simple: it involves only combining two easily computable quantities, the numbers of singletons and doubletons. Thus, this methods is available to colleagues without training in formal methods.
Third, Unseen Species and similar models for estimating missing data can be useful proxies in assessing whether and where usually scarce resources should be put to use. Their estimates can differ significantly from other measures of dataset sparsity, as we illustrate by comparing Chao1 coverage upper bounds and TTR.
Assessing, for instance, the present coverage of a database to be expanded can be a valuable piece of information when trying to obtain funding for this effort.

\section{Conclusions}\label{sec:conclusion}

Our main goal in applying Unseen Species models in these case studies was to demonstrate that they can be useful additions to the methodological repertoire of computational musicologists. Surely, it will not be hard to think of other areas where one could use this model.
We encourage our colleagues to engage with this kind of modeling in their own domains. However, we want to emphasise that, as the estimator rely on specific assumptions that may not hold for all scenarios, caution has to be applied regarding the validity of the conclusions to be drawn. For example, we are aware that the coverage percentages estimated in this article may in reality lie far from the true (but possibly unknowable) achieved coverage. The strength of the methodology, however, lies in the fact that it yields a upper bound for this quantity: there is at least that much to discover. 
In the end, our work is meant as an invitation to constructive criticism, which is enabled by the explicit nature of the approach. Computational modeling and critical thinking are not opposed (as sometimes suggested), but rather are the same thing in different disguise. 

\section*{Acknowledgements}

This work was supported by the Social Sciences and Humanities Research Council of Canada by the grant no. 895-2023-1002, Digital Analysis of Chant Transmission, and the project ``Human-centred AI for a Sustainable and Adaptive Society'' (reg. no.: CZ.02.01.01/00/ 23\_025/0008691), co-funded by the European Union.

\printbibliography

\clearpage
\appendix
\section{Tables}\label{appdx:tables}

\begin{table}[htbp]
    \centering
    \scalebox{.91}{
    \begin{tabular}{lrrrrrr}
\toprule
\textbf{Composer} & \textbf{Types} & \textbf{Tokens} & \textbf{TTR} & \bm{$f_1$} & \bm{$f_2$} & \textbf{Coverage} \\
\midrule
Leopold Koželuch (1747--1818) & 361 & 16598 & 0.022 & 77 & 74 & 0.900 \\
Wolfgang Amadeus Mozart (1756--1791) & 466 & 15272 & 0.031 & 157 & 82 & 0.756 \\
Ludwig van Beethoven (1770--1827) & 1722 & 50052 & 0.034 & 732 & 301 & 0.659 \\
Arcangelo Corelli (1653--1713) & 490 & 14314 & 0.034 & 191 & 66 & 0.639 \\
François Couperin (1668--1733) & 333 & 9472 & 0.035 & 147 & 40 & 0.552 \\
Heinrich Schütz (1585--1672) & 471 & 11709 & 0.040 & 161 & 74 & 0.729 \\
Franz Schubert (1797--1828) & 308 & 6200 & 0.050 & 0 & 71 & 1.000 \\
Johann Sebastian Bach (1685--1750) & 931 & 18493 & 0.050 & 390 & 143 & 0.636 \\
Domenico Scarlatti (1685--1757) & 733 & 12490 & 0.059 & 275 & 153 & 0.748 \\
Carl Philipp Emanuel Bach (1714--1788) & 698 & 11191 & 0.062 & 290 & 116 & 0.658 \\
Johann Christian Bach (1735--1782) & 314 & 5063 & 0.062 & 132 & 53 & 0.656 \\
Claudio Monteverdi (1567--1643) & 232 & 3289 & 0.071 & 111 & 38 & 0.589 \\
Felix Mendelssohn (1809--1847) & 1094 & 14758 & 0.074 & 448 & 181 & 0.664 \\
Frédéric Chopin (1810--1849) & 726 & 9125 & 0.080 & 226 & 137 & 0.796 \\
Pyotr Ilyich Tchaikovsky (1840--1893) & 278 & 3059 & 0.091 & 52 & 67 & 0.932 \\
Girolamo Frescobaldi (1583--1643) & 536 & 5318 & 0.101 & 248 & 85 & 0.597 \\
Jacopo Peri (1561--1633) & 316 & 2884 & 0.110 & 151 & 57 & 0.612 \\
Ignaz Pleyel (1757--1831) & 179 & 1567 & 0.114 & 67 & 44 & 0.778 \\
Antonín Dvořák  (1841--1904) & 177 & 1539 & 0.115 & 53 & 47 & 0.856 \\
Giovanni Battista Pergolesi (1710--1836) & 141 & 1189 & 0.119 & 58 & 16 & 0.573 \\
Edvard Grieg (1843--1907) & 1038 & 8236 & 0.126 & 193 & 340 & 0.950 \\
Georg Friedrich Händel (1685--1759) & 44 & 350 & 0.126 & 9 & 12 & 0.929 \\
Robert Schumann (1810--1856) & 265 & 1840 & 0.144 & 105 & 52 & 0.714 \\
Franz Liszt (1811--1886) & 755 & 5070 & 0.149 & 324 & 161 & 0.698 \\
Jan Pieterszoon Sweelinck (1562--1621) & 86 & 501 & 0.172 & 37 & 15 & 0.653 \\
Wilhelm Friedemann Bach (1710--1784) & 314 & 1753 & 0.179 & 158 & 56 & 0.585 \\
Clara Schumann (1819--1896) & 247 & 1326 & 0.186 & 99 & 43 & 0.684 \\
Nikolai Medtner (1880--1951) & 1332 & 6508 & 0.205 & 669 & 257 & 0.605 \\
Francis Poulenc (1899--1963) & 77 & 278 & 0.277 & 18 & 28 & 0.930 \\
Richard Wagner (1813--1883) & 402 & 1433 & 0.281 & 224 & 50 & 0.445 \\
Claude Debussy (1862--1918) & 291 & 1013 & 0.287 & 120 & 65 & 0.724 \\
Maurice Ravel (1875--1937) & 276 & 861 & 0.321 & 113 & 64 & 0.735 \\
Gustav Mahler (1860--1911) & 219 & 595 & 0.368 & 129 & 42 & 0.525 \\
Sergei Rachmaninoff (1873--1943) & 456 & 1141 & 0.400 & 280 & 87 & 0.503 \\
Erwin Schulhoff (1894--1942) & 251 & 488 & 0.514 & 137 & 61 & 0.620 \\
Béla Bartók (1881--1945) & 709 & 1191 & 0.595 & 513 & 104 & 0.359 \\
\midrule
\textbf{Total} & \textbf{6015} & \textbf{246166} & \textbf{0.024} & \textbf{2488} & \textbf{1097} & \textbf{0.681} \\
\bottomrule
\end{tabular}
    }
    \caption{Harmonic vocabularies of composers represented in the \textit{Distant Listening Corpus}, sorted by estimated coverage.}
    \label{tab:composers}
\end{table}

\end{document}